# Analysis of interaction partners of H4 histone by a new proteomics approach.

Evelyne Saade[1*], Undine Mechold[1¥], Arman Kulyyassov[*¤], Damien Vertut[*], Marc Lipinski[*] and Vasily Ogryzko[*]

[*] Institut Gustave Roussy, 39 rue Camille Desmoulins, 94805 Villejuif cedex, France

[¥] Institut Pasteur, URA 2171, Unite de Génétique des Génomes Bactériens, 75724 Paris cedex 15, France

[¤] National Center for Biotechnology, Valikhanov street, 43, Astana; Kazakhstan

[1] Authors have contributed equally to this work.

Corresponding author: Vasily Ogryzko, Institut Gustave Roussy, PRI, 39 rue Camille Desmoulins, 94805 Villejuif cedex, France. Email: vogryzko@gmail.com; Tel: +33142116525; Fax: +33142115494, http://sites.google.com/site/vasilyogryzko/

**ABSTRACT**

We describe a modification of the TAP method for purification and analysis of multiprotein complexes, termed here DEF-TAP (for Differential Elution Fractionation after Tandem Affinity Purification). Its essential new feature is the use for last purification step of 6XHis-Ni$^{++}$ interaction, which is resistant to a variety of harsh washing conditions, including high ionic strength and presence of organic solvents. This allows us to use various fractionation schemes before the protease digestion, which is expected to improve the coverage of the analysed protein mixture and also to provide an additional insight into the structure of the purified macromolecular complex and the nature of protein-protein interactions involved. We illustrate our new approach by analysis of soluble nuclear complexes containing histone H4 purified from HeLa cells. In particular, we observed different fractionation patterns of HAT1 and RbAp46 proteins as compared to RbAp48 protein, all identified as interaction partners of H4 histone. In addition, we report all components of the licensing MCM2-7 complex and the apoptosis-related DAXX protein among the interaction partners of the soluble H4. Finally, we show that HAT1 requires N-terminal tail of H4 for its stable association with this histone.

## INTRODUCTION

Identification of interaction partners of a protein of interest is an established way to get an insight into its function. The analysis of multiprotein complexes purified directly from the cells provides the most physiologically relevant information about the stable protein-protein interactions in vivo. In the laboratory of Y. Nakatani a system was established to search for interaction partners of a protein of interest, designed specifically for mammalian cells. It is based on generation, via retroviral vectors pOZ.FHN/C, of mammalian cell lines expressing a fusion of the protein of interest with a double epitope tag FLAG-HA, tandem-affinity purification (TAP) of protein complexes and their identification by mass spectrometry [1]. In this paper, we describe a modification of the pOZ vector system, which adds a 6XHis tag purification as an alternative option for the second purification step. Several well known features make $6XHis-Ni^{++}$ binding attractive as a last step in TAP protocol: 1) Elution from the $Ni^{++}$ agarose requires imidazole, which is easy to remove before the protease digest and LC-MS/MS analysis. 2) There is a wide variety of commercially available materials for purification of the His-tagged proteins. However, another feature of $6XHis-Ni^{++}$ interaction has been rarely exploited - its resistance to a number of harsh conditions, such as high ionic strength and presence of organic solvents, which usually are detrimental for the interaction of epitope-antibody interaction. One can take advantage of this property to fractionate the multiprotein complex after its immobilization on the $Ni^{++}$ agarose and before final elution step and protease digest. In addition to introducing a new orthogonal dimension to the separation procedure, fractionation of native proteins before protease digest should provide additional clues about the nature of protein-protein interactions in the analyzed multiprotein complex. In this paper we test feasibility of this approach, which we term here DEF-TAP (for Differential Elution Fractionation after Tandem Affinity Purification).

As a model system for validation of our approach, we have analyzed protein complexes that deposit H4 histone on chromatin. Chromatin is the hierarchically organized complex of DNA, histones and nonhistone proteins which is increasingly recognized as the principal carrier of epigenetic information [2], an information required to specify the state of a cell that cannot be deduced from genetic sequence. Most of the epigenetic information associated with chromatin is believed to be encoded in core histones, the protein components of the basic unit of chromatin, nucleosome [3]. The candidate epigenetic marks associated with histones are their post-translational modifications and replacement variant histones [4]. The mechanism of chromatin replication is one of the subjects of molecular epigenetics, an emerging discipline concerned with the molecular mechanisms of functioning of epigenetic information.

Chromatin replication involves both transfer of parental histones to new double stranded DNA and the assembly of new nucleosomes de novo. Recent research has uncovered a diversity of protein chaperones involved in the deposition of histones on eukaryotic DNA [5-8]. The proteomic analysis of different proteins associated with soluble histone H4 is expected to help in understanding different aspects of H4 deposition on chromatin.

New H4 is acetylated prior to its deposition onto DNA and is subsequently deacetylated following nucleosome assembly [9]. Although inhibiting deacetylation of the incorporated H4 prevents the complete maturation of newly replicated chromatin, the function of the acetylation of H4 prior to its deposition is unknown [10, 11]. One histone acetyltransferase (HAT) that catalyzes the acetylation of newly synthesized H4 on lysines 5 and 12 is the "type B" HAT1 histone acetyltransferase that has been conserved throughout evolution [12]. HAT1 functions as a part of a complex, which in humans also contains RBAp46 [13]. HAT1 enzymes from other species contain orthologs of p46 [14, 15]. We focus on H4 interaction with these proteins as a model to test the feasibility of our approach to analyze structure and composition of multiprotein complexes.

**MATERIALS AND METHODS:**

**Recombinant DNA**

Vector pOZFHHC expressing the C-terminally tagged histone H4 (pOZFHHC.H4) was made by introducing H4 ORF, amplified from the EST clone AI589002 (Invitrogen) between the XhoI and NotI sites of the pOZFHHC vector in frame with the FLAG-HA-6XHis tag peptide sequence. The pOZFHHC.H4T vectors expressing H4 histone lacking the N-terminal peptide MSGCGKGGKGLGKGGAK was similarly generated; the N-terminal peptide sequence in this case starts with MGRHRKVL.

**Immunology**

For the Western analysis, nuclear extracts or complex preparations were separated on SDS-PAGE. Proteins were transferred to nitrocellulose membranes and probed with antibodies according to the standard procedure [16].

Antibodies against mcm2 (ab 4461), mcm3 (ab 4460), mcm5 (ab 17967), DAXX (ab 2017), RbAp46 (ab 3535) and histone H4 acetyl K5 (ab 51997) were from Abcam; antiHA antibody (clone 3F10) was from Roche; acetyl-histone H4 Lys12 antibody (ab 2591S) was from Cell Signaling, Histone H4 polyclonal antibody (ab 3624) was from BioVision. The remaining antibodies against the MCM2-7 proteins were a generous gift of Dr. Rolf Knippers (Department of Biology, Universität Konstanz, Germany).

For immunofluorescent staining, the e:H4 and e:H4T expressing Hela cells were fixed with 4% Paraformaldehyde, permeabilized with 0.3% Triton and blocked with 0.3% BSA before treatment with anti-HA and secondary antibodies. The confocal images were taken on the Leica TCS SP equipped with the argon and UV lasers.

**Cell culture and cell lines**

HeLa S3 were grown in DMEM with 10% FBS. The HeLa S3 expressing the epitope-tagged histones were generated by retroviral transduction and magnetic sorting according to [1].

**Biochemistry**

For chromatin isolation, nuclei were prepared by incubating cells in hypotonic buffer (10mM Tris HCl pH 7.5, 10mM KCl, 1.5mM $MgCl_2$, 0.2mM PMSF, 15mM β-mercaptoethanol, 0.1% Triton X100), then they were resuspended in sucrose buffer (0.34M Sucrose, 10mM Tris HCl pH 8, 3mM $MgCl_2$, 15mM β-mercaptoethanol, 0.2mM PMSF) and treated with micrococcal nuclease (Sigma). After stopping the reaction with 10mM EDTA, the nuclear debris was removed by centrifugation and the digested chromatin was fractionated on the 10-35% glycerol gradient (with the buffer composition 20mM Tris HCl, 0.2 mM EDTA, 300 mM NaCl, 0.2mM PMSF) for 5 hours at 40,000 g using benchtop ultracentrifuge (Beckman Optima TL).

Purification of the e:H4 complexes from the nuclear extracts was essentially done as in [1] except for the modification at the last step: the FLAG peptide eluate from the first purification step was added to Ni-NTA agarose (QIAGEN) and incubated for 2 hours at 4˚C. The Ni-NTA beads were washed 5 times with 20 mM Tris pH 8.0, 10% Glycerol, 0,1% Tween, 20mM imidazole. The final elution step was done with 250 mM imidazole in the same buffer. For the shotgun analysis, the Tween was omitted in the elution buffer and for the two last washing steps. The imidazole was removed and the proteins concentrated and denatured by phenol-ether extraction [17].

**HAT assay**

The HAT assay mix was 2 µl of 10Xbuffer A (500 mM Tris-HCl, pH 8.0; 10 mM DTT; 1 mM EDTA), 10% glycerol (2 µl), 1 µl of histones (0.2µg), 1 µl of H4 complex, 0.2 mM of PMSF, Na-Butyrate (10 mM final) and C14 acetylCoA- 0.1 µl (25 nCi/assay, 2-10 Ci/ mMol, Amersham). Samples were incubated for different time and the reaction was terminated by adding 5 µl of SDS page sample buffer 5X (200 mM Tris-HCl pH 6.8, 5% SDS, 50% Glycerol) + ß-mercaptoethanol 10%. After, histones were separated on an 18% acrylamide SDS page gel, the gel was fixed, stained with coomassie, dried and incorporation of radioactive acetyl CoA was followed using phosphoImager (Fujifilm FLA-3000).

**Mass-spectrometry analysis**

The proteins bands were excised from the gel and processed as in [18]. Alternatively, for the shotgun approach, phenol/ether extraction [17] was used to remove the imidazole and concentrate and denature the proteins. After, ammonium bicarbonate was added to final concentration of 25mM, and trypsin digestion was performed overnight (12.5ng/ul). The peptide mixtures obtained either from tryptic digestion of the bands on the SDS-PAGE or from direct digestion of protein complexes, were analyzed by nano-HPLC (LC Packings) coupled to ion trap mass spectrometer (ThermoFinnigan LCQ Deca XP) equipped with a nanoelectrospray source. The reversed-phase HPLC separation was performed with the flow rate 300 nl/min using a Nano C18 PepMap 100 pre-column (5 mm, 100 Å, 300 µm I.D. × 1 mm), coupled with a column of 75 µm

I.D. × 15 cm with the same resin (LC Packings, Dionex, Voisins le Bretonneux, France), with a linear gradient of $H_2O$:ACN from 3% ACN to 55% ACN. The samples were run in two different modes. For peptide identification, the ion trap acquired successive sets of 6 scan modes consisting of: full scan MS over the ranges of 200-2000 m/z, followed by 5 data-dependent MS/MS scans on the 5 most abundant ions in the full scan. The MS/MS spectra were acquired with relative collision energy of 35% and an isolation width of 2.0 Daltons. Their interpretation was performed with the Bioworks software package. All proteins reported had at least one peptide with an identification score (expressed as Xcorr value) of 2.5 or higher. Alternatively, for the confirmation and quantification of the presence of a particular peptide in the sample, the ion trap was set in a MRM mode. The sample was separated using the same nanoLC gradient, and the ion trap was set to isolate, fragment and MS/MS scan of 5 parental ions having predetermined M/Z ratios. The relative collision energy was 35% and an isolation width was 4.0 Daltons. The peptide identity was confirmed by comparison of the obtained MS/MS spectra with the previously acquired spectra for the same peptide from the identification analysis. The relative quantity of every peptide in different fractions was estimated by comparison of the peak areas in the Total Ion Chromatograms (TIC) for this peptide obtained from MRM analysis of these fractions.

**RESULTS**

**Purification of soluble H4 complexes from HeLa cells using new pOZFHHN/C vectors**

The novel vectors pOZFHHN and C are derivatives of the pOZFH vectors described previously [1]. In addition to FLAG and HA tags, they also contain 6XHis tag. (Fig. 1A) This design allows one to perform standard tandem FLAG-HA purification, and also to use 6XHis tag instead of HA step. To test how our approach works, we generated HeLa cell line expressing H4 histone tagged on its C-terminus (e:H4). According to several criteria, the e:H4 was incorporated into chromatin. First, e:H4 co-sedimented with DNA during glycerol gradient fractionation of chromatin, prepared from the e:H4 expressing cells and partially digested with micrococcal nuclease. Fig.1B illustrates the presence of tagged histones (left, bottom) and DNA (left, top) in the same fractions, suggesting that the tagged histones are in a complex with DNA. Second, staining of a cellular population with anti-HA antibodies showed that e:H4 was present in the mitotic chromosomes (figure 1B, right panel). Further immunofluorescence experiments were used to confirm that the association of tagged H4 with DNA is specific. Given that the histones incorporated into nucleosomes resist modest salt treatments, the e:H4 expressing cells were permeabilized and washed with buffer containing up to 600mM NaCl before fixation. Epitope tagged e:H4 was found in the pellet fraction after salt extraction (not shown). Therefore, we concluded that e:H4 histone was specifically incorporated into chromatin.

Nuclei extracts were prepared from the e:H4 expressing cells and subjected to modified TAP procedure using as a first step, anti-FLAG purification, and as a second step, 6XHis purification followed by imidazole elution (Materials and Methods). As H4 extracted with 0.4M KCl is not incorporated into chromatin, the identification of proteins associated with this form of H4 histone should provide clues on the mechanisms of its deposition and related processes.

The composition of the purified e:H4 complexes was analyzed by SDS-PAGE. As seen from the figure 1C (lane H4), numerous bands are coeluted by imidazole with epitope tagged H4 at the last stage of purification. Importantly, mock purification from the parental cell line done in parallel did not yield any significant material.

Some of the protein bands were excised and the proteins were identified by mass spectrometry. The 48 kDa band was identified as a mixture of RbAp46 and RbAp48 proteins.

The 50 kDa band was identified as HAT1. Two bands at the bottom of the gel were identified as histone H4 and histone H3.

Notably, endogenous H4 was nor present in our preparation as tested by silver staining, or using Western blotting with antiH4 antibody (Fig. 1C, right panel), indicating that in our preparation H4 exists as a dimer with H3 histone. Thus, our analysis is consistent with previous reports that H4 is deposited on chromatin as a dimer with H3 [8].

**Shotgun analysis identifies histone chaperones and replication licensing factors among the H4 interaction partners**

To test the compatibility of the shotgun approach with our modified procedure, the purified H4 complexes were digested with trypsin directly after imidazole removal and denaturation (see Materials and Methods). The resulting peptide mixture was directly analyzed with LC-MS/MS. The proteins found by this method are listed in the Table 1. In addition to the previously identified H4, H3, HAT1, RbAp46 and RbAp48, we found several known histone and protein chaperones in the mixture of the H4 associated peptides. This finding was expected, as most of the H4 that can be extracted from nucleus with 400 mM salt is not incorporated into chromatin and most likely is present in the nucleus as part of deposition complexes. A somewhat unexpected finding was DAXX, an apoptosis associated protein [19]. Its presence among H4 interacting proteins was confirmed by Western analysis (Fig. 2A).

Other findings were more unexpected. We also identified four proteins belonging to the replication licensing complex: Mcm2, Mcm4, Mcm6 and Mcm7. Western analysis confirmed their presence in the H4 complex. The proteins are subunits of a larger Mcm2-7 DNA helicase complex, believed to play a role of a licensing factor in eukaryotic DNA replication. Although other subuntis of the Mcm2-7 complex have not been detected by LC-MS/MS, our observation prompted us to analyze the same samples with antibodies specific for all components of Mcm2-7 complex. Indeed, all proteins were readily observed by Western blotting (Fig.2A). For several MCM proteins, the interaction with H4 histone was further confirmed by reciprocal co-immunoprecipation (Fig. 2B).

**HAT1 and RbAp46 but not RbAp48 require H4 N-terminal tail for stable association with H4**

Identification of various interaction partners of H4 prompted us to ask whether direct shotgun analysis after TAP purification could be used to compare the composition of different multiprotein complexes. As a model for our studies, we choose to delete the N-terminal tail of H4 and identify proteins that were interacting with the truncated H4 (H4T, for tailless H4), as we expected that the association of some proteins with soluble H4 might depend on the presence of its N-terminal tail. A cell line expressing H4 lacking aa 1 to 17 (MSGCGKGGKGLGKGGAK) was therefore generated. Using the same criteria as above (association with DNA on glycerol gradient, association with mitotic chromosomes), H4T was incorporated into chromatin as its full length counterpart (data not shown). To further compare the efficiency of incorporation of e:H4 and e:H4T into chromatin, nuclei were extracted from e:H4 and e:H4T expressing cells using 1.5 M NaCl. Western blotting with αHA antibodies revealed no difference in the distribution between soluble and pellet fractions (Fig. 3A).

The e:H4T and e:H4 complexes (Fig. 3B) were directly analyzed by tryptic digestion of the imidazole eluate using LC-MS/MS. Interestingly, the e:H4T sample did not contain any peptides corresponding to the HAT1 and RbAp46 proteins, otherwise easily found among the interaction partners of the full length H4. Other proteins, including RbAp48 closely related to RbAp46, were still detectable in both preparations. Two peptides, RLNVWDLSK and TPSSDVLVFDYTK, present in both RbAp46 and RbAp48, were found in smaller amounts in the e:H4T sample than in e:H4. The difference observed between the compositions of H4 and H4T complexes was further confirmed by MRM analysis using the same mass-spectrometer (Supplemental Fig.1B) and by Western blotting with antibodies directed specifically against HAT1 (Supplemental Fig.1A) and RbAp46 (Fig. 3C). The loss of RbAp46 in the e:H4T complex could explain the significant decrease in the amounts of the RbAp46/RbAp48 band in this complex as compared to the e:H4 complex (Fig. 3B).

The absence of HAT1 protein among H4T associated proteins prompted us to compare the HAT activity in the e:H4 and e:H4T complexes. In the HAT assay, using $C^{14}$-labeled AcetylCoA shown on in figure 3D, exogenous histone H4 (and histone H2A to some extent) was found acetylated by the e:H4 complex. This contrasted with the e:H4T complex, where no acetylation activity could be detected. Thus, consistent with the role of the N-terminal tail of H4 in the recruitment of HAT1/RbAp46, the tail is responsible for stable association of the soluble H4 histone with a major histone acetytransferase activity.

**The nature of H4 interaction with HAT1 and RbAp46 is different from that of HAT1-RbAp48**

6XHis-$Ni^{2+}$ interaction is known to be resistant to various harsh washing conditions, in particular, to high ionic strength and organic solvents. We took advantage of this property by using a step gradient of acetonitrile in order to progressively elute the interaction partners of the immobilized e:H4. This was followed by 300 mM imidazole to elute e:H4 with remaining proteins.

As expected, H4 specific peptides were detected by MRM in the imidazole eluate only, consistent with the resistance of $Ni^{++}$-6XHis interaction to acetonitrile. HAT1 and RbAp46 eluted with acetonitrile very similarly, with the majority of the two proteins detected in the acetonitrile fractions 2 and 3 (Fig. 4A). Intriguingly, the RbAp48 protein showed a different fractionation profile, as it was mostly detected in the imidazole eluate. The distribution of these proteins in different fractions was confirmed by Western analysis (Fig. 4B).

The e:H4 complex, immobilized on $Ni^{++}$-NTA beads was also fractionated with a step-gradient of NaCl (600 mM, 800 mM, 1M, 1.4M, 1.7M, 2M, and 300 mM imidazole). As in the previous experiment, H4 was detected only in the imidazole fraction. Importantly, pRbAp48 eluted more readily in 600/800 mM NaCl (78%) compared to RbAp46 and HAT1 (70%) (not shown). Thus, although less pronounced, the difference between fractionation of HAT1 and RbAp46 on one hand, and RbAp48 on the other, could also be seen in this second fractionation procedure.

Taken together, these findings indicate that the nature of the interactions of HAT1/RbAp46 with H4 histone is different from that of RbAp48. This is consistent with our observation that the amino-terminal tail of H4 is required only for the binding to the HAT1 and RbAp46.

**H4 is acetylated when in a complex with HAT1**

The procedure of purification of multi-protein complexes detects interactions that are resistant to the extraction and other biochemical manipulations. Given the fact that the N-terminus of H4 is the target of HAT1-directed acetylation, our finding that the N-terminal tail of H4 is required for stable binding to HAT1 was somewhat surprising, as one expects a rather transient interaction between an enzyme and its substrate. To better understand the nature of the interaction between HAT1 and H4, we tested whether H4 is acetylated when present in a complex with HAT1. Cells expressing epitope-tagged HAT1 were generated, and the proteins in the complex with e:HAT1 were purified by FLAG affinity purification. The presence of H4 and its acetylation status was analyzed by Western blotting. The acetylation of lysines 5 and 12 of H4, the known deposition-related modifications of H4, were readily observed in the e:HAT1 pulldown (Fig. 5A). In fact, H4 bound to HAT1 appeared to be enriched in these modifications, as seen from comparison with the H4 that was not bound to HAT1 (flowthrough fraction). We thus conclude that the acetylated form of H4 is present in the HAT1 complex. Intriguingly, we also observed a relative increase in the levels of H4AcK5 and H4AcK12 in the cell lines expressing e:HAT1 (Figure 5, compare NE column for C (control) and HAT1 samples), consistent with the notion that HAT1 is responsible for these modifications of H4.

**DISCUSSION**

**Methodology**

We have presented here a modification of the TAP method for purification and analysis of multiprotein complexes, termed DEF-TAP (for Differential Elution Fractionation after Tandem Affinity Purification). The newest feature in this approach relies on the properties of the 6XHis interaction with $Ni^{++}$, that allow for harsh washing conditions, inappropriate in the case of regular antibody-epitope interactions. Various fractionation schemes (i.e., step gradients of salt or organic solvent such as acetonitrile) can then be applied to the purified complex while still immobilized on the solid support via 6XHis-$Ni^{++}$ binding. The benefits of protocol are two-fold: 1) It adds a new orthogonal dimension to the repertoire of separation techniques used to reduce complexity of analyzed samples, 2) The fractionation of native proteins before protease digest can provide additional clues about the substructure of the analyzed complex, as well as about the physicochemical nature of protein-protein interactions therein (given that different kinds of intermolecular interactions are sensitive to different types of treatments). In addition to organic solvents and high ionic strength, 6XHis-Ni-NTA interaction is also resistant to Guanidine HCl (up to 6M), urea (8M), Triton (up to 2%), SDS (up to 0.3%), glycerol (50%) and ethanol (20%), which can also be used as alternative methods of fractionation. In summary, our modification extends the arsenal of existing protocols for analysis of protein-protein interactions in vivo.

Here, we have applied this approach to provide further insight into the structure of the multi-protein complexes containing histone H4. The results of both acetonitrile and salt fractionations points to similarities in the interaction between H4 on one hand and HAT1 and RbAp46 on the other hand, and also suggest that the RbAp48 interacts with H4 in a different manner. Given the high degree of similarity between RbAp46 and RbAp48, this observation was rather unexpected and most likely indicates that additional proteins could play a role in the stable interactions of RbAp46 and RbAp48 with histone H4 (such as HAT1 in the case of RbAp46). The difference in the fractionation behavior of the two proteins is consistent with our observation that the N-terminal tail of H4 is required for interaction with the HAT1/RbAp46, but not with RbAp48. Taken together, our findings support the notion that HAT1/RbAp46 form a structural unit and most likely a separate complex with H4 histone, consistent with previous data obtained with studies in other organisms (reviewed in [20]), and validating the use of the DEF-TAP protocol to study multiprotein complexes. We have also noticed that, despite overall similarity in the fractionation profiles of HAT1 and RbAp46, HAT1 requires a higher acetonitrile concentration for elution (Fig. 4). This slight difference in behavior might reflect the fact that hydrophobic interactions are more important for the RbAp46 interaction with H4 than for the HAT1-H4 interaction. In this regard, it is of note that RbAp46 also interacts with H3, and that the H3-H4 dimer is stabilized mostly by hydrophobic interactions.

**Licensing factors in the soluble H4 complexes**

The identities of the proteins bound to soluble H4 are in good agreement with the notion that soluble H4 molecules are ready to be incorporated in chromatin and thus require chaperones. An unexpected finding was the presence of MCM2-7 proteins H4 complex. The MCM2-7 complex is believed to serve as a licensing factor that marks unreplicated origins prior to being released upon their replication, thus ensuring that each origin fires only once during S-phase. The interaction of MCM2-7 complex, specifically via MCM2, with histone H3 has been already demonstrated to occur with high efficiency *in vitro* [21]. However, whether this had any *in vivo* significance was not clear. Moreover, it was proposed that the detected *in vitro* interaction reflected an interaction with chromatin, implying that the *in vivo* interactor of the MCM complex would be nucleosomal H3 (although it was not clear how the H3 domain crucial for this interaction would be accessible). Our finding that MCM2-7 interacts with the soluble form of H4 is most likely due to the presence of stoichiometric amounts of H3 in the deposition ready H3/H4 dimers. Although consistent with previous findings, it sheds a completely new light on the possible relevance of the MCM-H3 interaction previously reported. Almouzni and coworkers [22] have proposed a model whereby MCM2-7 together with histones H3 and H4 and the histone chaperone ASF1 form an intermediate structure that tunes histone supply to chromatin and replication fork progression.

**Role of HAT1/RbAp46 and H4 tail acetylation in histone deposition**

Consistent with the previous reports [20], we have observed that H4 is acetylated in a complex with HAT1. Together with the finding that the N-terminal tail of H4, the substrate of HAT1, is essential for stable interaction with the HAT1/RbAp46 complex, this underscores somewhat paradoxical situation. Enzymes usually have relatively low affinity for their products, as the strong binding to the product prohibits efficiency of catalysis by slowing the turn-over of the enzyme. On the other hand, our ability to purify acetylated H4 in the complex with HAT1 suggests that this binding is strong enough to withstand many hours of biochemical separation. Further research is warranted by the unusual nature of the interaction between HAT1 and its substrate.

The first question is the structural basis for this interaction. It remains to be explored a) Whether the acetylated H4 tail continues to play a role in stability of the HAT1/H4 complex, for example, by remaining in the active center of the HAT enzyme, and b) What is the role of RbAp46 protein in this complex. Concerning the latter question, further studies will have to take into account that both RbAp46 and his close homologue RbAp48 can bind H4 directly, with residues 16-41 of histone H4 being sufficient for the recognition [13, 23]. Only Lysine 16 is absent in the truncated version of H4 (its sequence is MGRHRKVL…), and the fact that RbAp48 remains associated with H4T suggests that the this residue is not crucial for the binding. Nevertheless, the complete absence of HAT1 and RbAp46 from the complex with the truncated version of H4 lacking the N-terminal tail and very different fractionation behavior of RbAp46 compared to RbAp48 suggests that other interaction partners can significantly modulate interaction of these proteins with histone H4.

Regardless of the structural basis of the stable HAT1/H4 interaction and role of RbAp46, a different question is the biological significance of a stable complex between an enzyme and its product. On one hand, keeping acetylated H4 in a multiprotein complex might create a 'microcompartment' protecting H4 from the action of histone deacetylases, abundant both in the cytoplasm and in the nucleus. On the other hand, at the later stages of the deposition process, H3/H4 dimers are transferred to complexes that contain CAF1, HIRA and other histone chaperones, responsible for specific deposition pathways [8]. Why this transfer does not occur immediately after acetylation, and why a proportion of acetylated H4 has to be kept in the form of a HAT1/H4 complex? The strategy of 'keeping options open' rather than committing to a particular deposition pathway might prove more beneficial in the case when the exact demand for each pathway is uncertain and depends on many circumstances, such as need for the cell to proliferate or differentiate (Fig.5B).

## ACKNOWLEDGEMENTS

We thank Dr. Rolf Knippers (Department of Biology, Universität Konstanz, Germany) for his generous gift of anti MCM antibodies. The work was supported by by grants from "La Ligue Contre le Cancer" (9ADO1217/1B1-BIOCE), the "Institut National du Cancer" (247343/1B1-BIOCE) and the National Center for Biotechnology (Kazakhstan) to VO.

**FIGURES WITH LEGENDS**

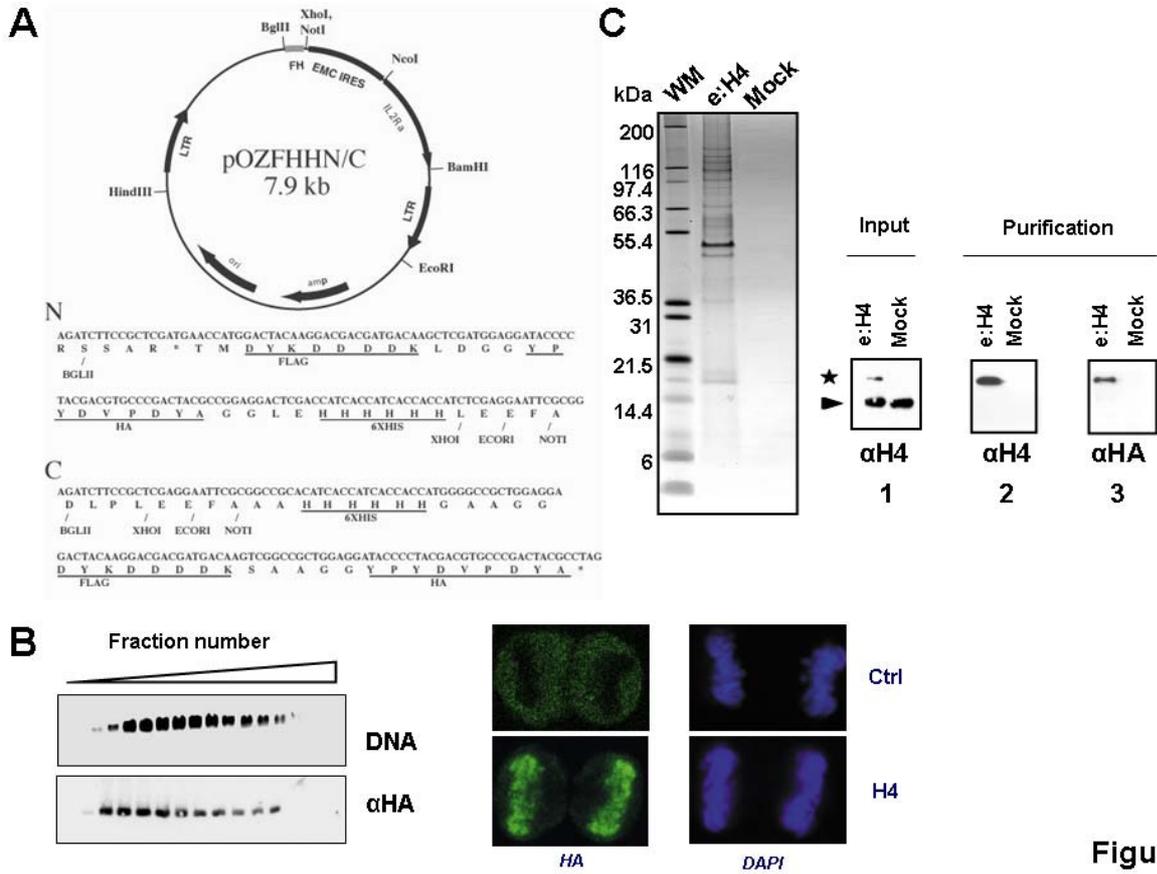

**Figure 1. Purification of e:H4 complex.**

A) **Retroviral pOZFHHN/C vectors**.

**Top panel.** Map of the pOZFHHN vector showing its principal components. ORF of the gene of interest is inserted between the XhoI and NotI restriction sites in frame with FHH sequence, encoding epitope-tag FLAG-HA-6XHis. Downstream of the ORF of the gene of interest, EMC IRES directs expression of ORF of IL2Rα, used as surface marker for selection.

**Bottom panel.** Nucleotide sequence of the region around the cloning site for both pOZ.FHHN (N) and pOZ.FHHC (C) vectors designed for the N- and C-terminal tagging, respectively. The amino acid sequences of the FLAG, the HA, and the 6XHis tags are underlined and the cloning sites are pointed out.

**B) e:H4 is incorporated into chromatin**.

**Left panel.** Chromatin was prepared from the e:H4 expressing cells and partially digested with micrococcal nuclease. It was fractionated on 10-35% glycerol gradient. 20 fractions were analysed on the presence of e:H4 by western blotting with HRP-conjugated αHA antibody (bottom). The presence of DNA was also monitored in the same fractions by agarose gel electrophoresis (top). A strong correlation in the presence of e:H4 (bottom) and DNA (top) in the same fractions strongly indicates that e:H4 is in a complex with DNA.

**Right panel**. Staining of mitotic chromosomes with αHA antibody and DAPI in control cells (Ctrl) and e:H4 expressing cells (H4).

**C) Purification of the e:H4 complex.**

**Left panel**. SDS page analysis and silver staining. WM – SeeBlue weight marker (Invitrogen), e:H4 – TAP purification of H4 complex, Mock - mock purification from the parental HeLa cell line not expressing e:H4. **Right panel.** Western blot analysis. 1 - αH4 antibody to detect H4 in extracts from the e:H4 expressing (e:H4) and parental cells (mock) prior to purification. 2 - αH4 antibody to detect the total H4 in the e:H4 and Mock samples after purification. 3 - αHA antibody to detect the tagged e:H4 in the e:H4 and Mock samples after purification.

The arrow points to the position of the endogenous H4. The asterisk indicates position of e:H4.

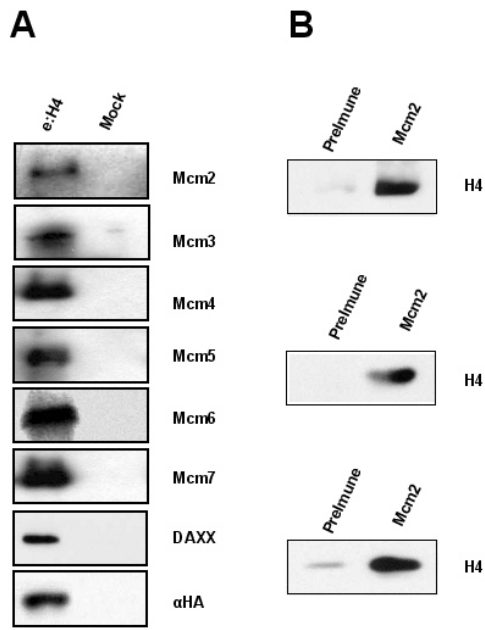

**Figure 2**

**Figure 2. MCM2-7 and DAXX proteins are interaction partners of e:H4.**

(A) Western blot detection of mcm and DAXX proteins in the e:H4 complex. The samples from the Figure 1C (e:H4, left; and mock purification, right) were separated on SDS-PAGE and probed with antibodies against the corresponding proteins (B) Coimmunoprecipitation of H4 histone with the mcm2, mcm3 and mcm5 proteins. The mcm proteins were immunoprecipitated from the Hela nuclear extracts using corresponding antibodies and the presence of H4 was detected by Western with αH4 antibody (left – preimmune serum, right – specific antibody).

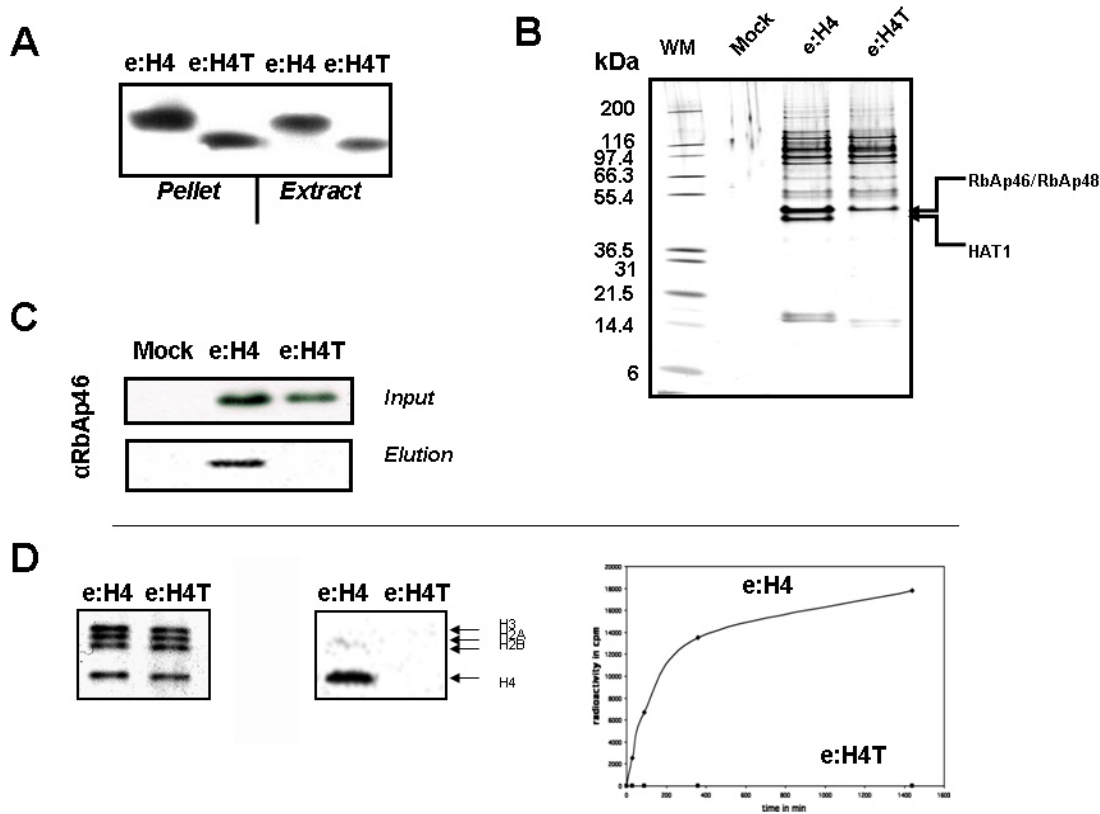

**Figure 3**

**Figure 3. Requirement of N-terminal tail of H4 for stable interaction with HAT1 and RbAp46.**

**(A) Similar degree of incorporation of e:H4 and e:H4T histones into chromatin.** Nuclei prepared from the e:H4 and e:H4T expressing cell lines were extracted with 1.5 NaCl, and the proportion of solubilized e:H4 and e:H4T (Extract) relative to the pellet fraction (Pellet) was compared by Western Blot analysis with αHA antibody.

**(B) Purification of the e:H4 and e:H4T complexes.** Silver stained SDS-PAGE gel shows proteins from mock purification (Mock), purification of e:H4 and e:H4T. WM - SeeBlue weight marker (Invitrogen). The arrows indicate HAT1 and RbAp46/RbAp48 bands.

**(C) RbAp46 is absent in H4T complex.** Western Blot with anti RbAp46 of input (top) and eluted (bottom) samples from e:H4 (middle) and e:H4T (right). As a control, mock purification was performed from HeLa cells not expressing an epitope-tagged proteins (left).

**(D) Loss of HAT activity in the e:H4T complex.** Histones were incubated with TAP-purified complexes from e:H4 and e:H4T sources in the presence of $C^{14}$ labeled AcetylCoA, resolved on SDS-PAGE gel, stained with coomassie and exposed for autoradiography.

**Left panel** shows equal presence of substrate histones as stained by coomassie .

**Middle panel** - autoradiography revealing acetylated bands only.

**Right panel** - the kinetic of radioactive incorporation for e:H4 and e:H4T complexes.

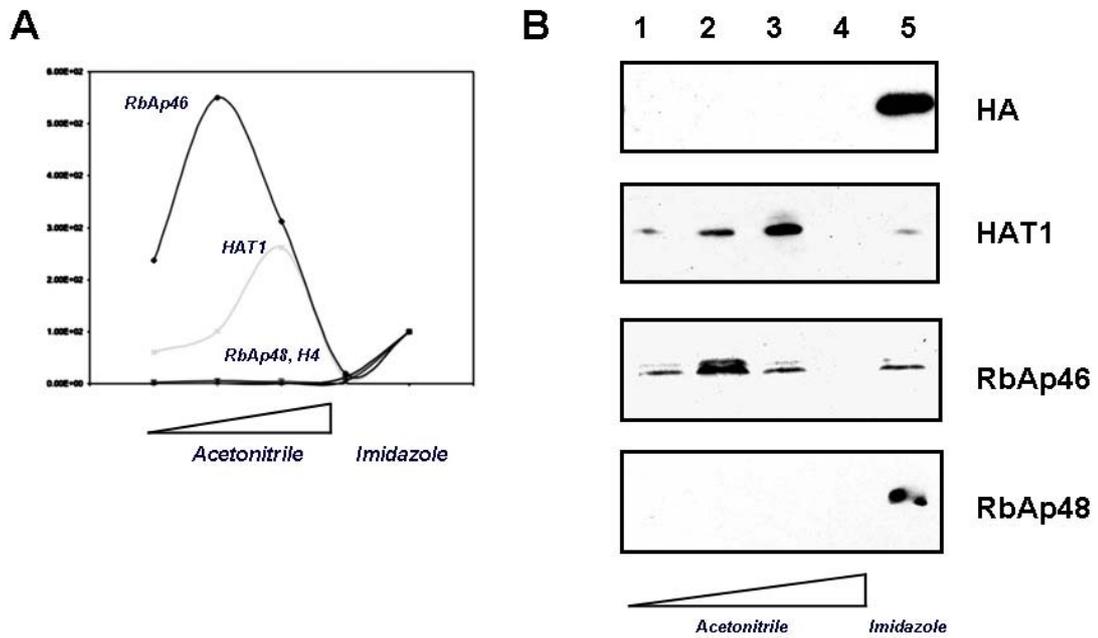

**Figure 4**

**Figure 4. Different nature of H4 interaction with HAT1/RbAp46 and RbAp48.**

(A) MRM analysis of the elution behaviour of the HAT1, H4, RbAp46 and RbAp48 proteins in the presence of an organic solvent. The e:H4 containing complex was purified as in legend to Figure 1C, except before the imidazole elution step, the $Ni^{++}$ agarose was consecutively incubated with increasing concentrations of acetonitrile (10%, 20%, 30%, 40%), followed by final 300 mM imidazole elution step.

(B). Western blot analysis of the same samples.

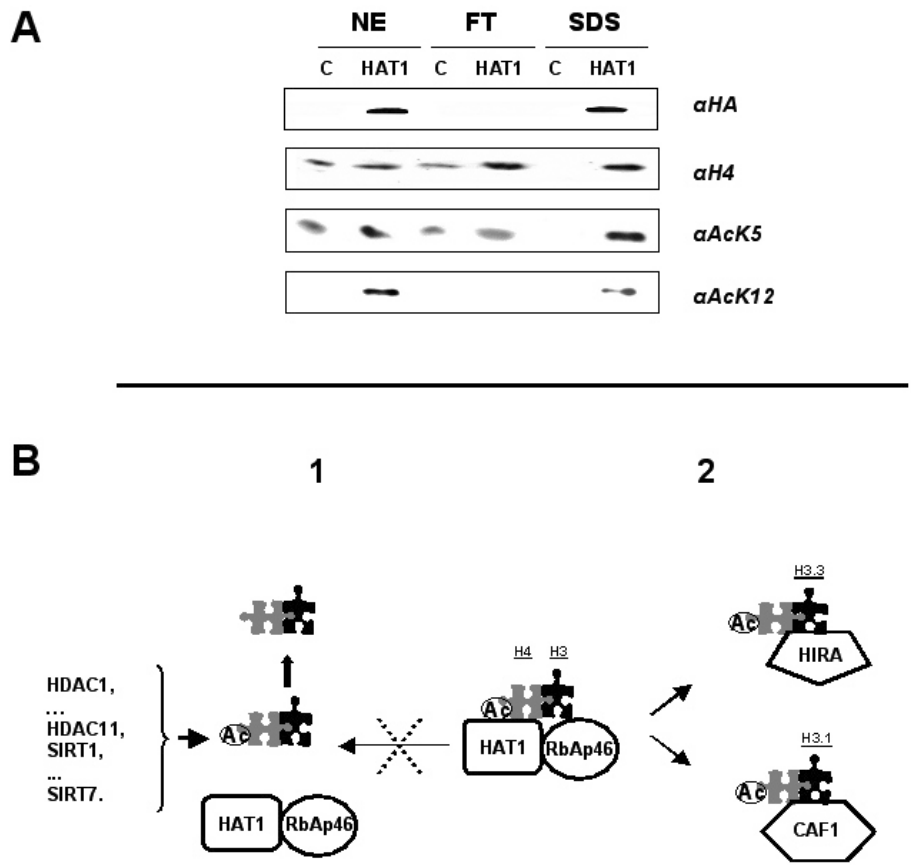

**Figure 5. Acetylation of H4 by HAT1.**

**A. Acetylated H4 present in HAT1 complex.**

Western analysis of the input nuclear extract (NE), flowthrough (FT) and SDS eluate (SDS) after FLAG affinity pulldown. The presence of HAT1 was assessed using αHA antibody. Presence of total histone H4 or its acetylated forms (on lysine 5 (AcK5) and lysine 12 (AcK12)) were detected by Western blot analysis with specific antibodies.

**B. Role of tight association of acetylated H4 with HAT1/RbAp46.**

(1) Dissociation of H4 from the HAT1/RbAp48 after its acetylation (Left) could lead to its rapid deacetylation due to the presence of numerous histone deacetylases (HDACs, SIRTs) in cell.

Tight association of H3/H4 dimer with HAT1/RbAp46 complex (Center) creates a 'microcompartment', avoiding the unnecessary deacetylation of H4.

(2) Premature transfer of the acetylated H4/H3 dimer to a particular histone chaperone (Right) will commit it to one deposition pathway and deprive alternative deposition pathways that might also require acetylated form of H4. Tight association of H3/H4 dimer with the HAT1/RbAp46 complex will 'keep options open', which could be more beneficial when the exact demand for each pathway is uncertain and depends on many circumstances, such as need for the cell to proliferate or differentiate.

| Name | Molecular Weight | Accession Number | Function | Detected peptides (*highest score; charge state*) |
|---|---|---|---|---|
| **Histone H4** | 11,4 | 4504301 | core histone | DNIQGITKPAIR (*3.5;2*); ISGLIYEETR (*3.8;2*) ; TVTAMDVVYALK (*3.5;2*); VFLENVIR (*1.8;1*); DNIQGITKPAIR (*3.2;2*); KTVTAMDVVYALK (*3.1;2*); |
| **Histone H3** | 15,4 | 4506439 | core histone | DIQLAR (*2.8;2*); STELLIR (*3.2;2*) |
| **FK506 binding 14** | 22 | 8923659 | protein chaperone | LSKDEVKAYLK (*4;2*) |
| **ASF1** | 23 | 7661592 | histone chaperone | NILASNPRVTR (*2.8;2*) |
| **DNAJ hom, C, memB** | 30 | 27597059 | protein chaperone | VSLEDLYNGK (*3.3;2*); DGNDLHMTYK (*2.6;2*) |
| **Nucleophosmin** | 32,5 | 10835063 | nucleolar phosphoprotein | MTDQEAIQDLWQWR (*3.9;2*) |
| **RbAp46** | 48 | 4506439 | HAT1 and HDACs interactor | EMFEDTVEER (*2.9;2*); TPSSDVLVFDYTK (*4.4;2*); RLNVWDLSK (*3.1;2*); YMPQNPHIIATK (*2.6;2*); LMIWDTR(*2.9;2*); TVALWDLR (*2.8;2*) |
| **RbAp48** | 48 | 5032027 | CAF1 subunit | EAAFDDAVEER; TPSSDVLVFDYTK(*4.4;2*); RLNVWDLSK(*3.1;2*); LNVWDLSK (*3.1;2*); LMIWDTR (*2.9;2*); TVALWDLR (*2.8;2*) |
| **HAT1** | 49,5 | 4504341 | H4 tail acetylation | FPEDLENDIR(2.9;2); LLVTDMSDAEQYR (*4.9;2*); WHYFLVEYK (*2.2;2*) |
| **DAXX1** | 81 | 4503257 | apoptosis | QTGSGPLGNSYVER (*2.7;2*); LEQLLALYVAEIR (*4;2*) SPMSSLQISNEK (*3.4;2*); QQLQLMAQDAFR (*2.1;2*) |
| **NASP1** | 85 | 477321 | H1 chaperone, in H3 complex | EAQLYAAQAHLK (*3;2*); EVSEEQPVVTLEK (*2.6;2*); EIEELKELLPEIR (*2.9;2*); SLLELAR (*3.1;2*) |
| **Mcm 6** | 92,8 | 7427519 | licensing factor | IQETQAELPR (*3.5;2*) |
| **Mcm4** | 96 | 1705520 | licensing factor | ALADDDFLTVTGK (*2.9;2*); DYIAYAHSTIMPR (*3.1;2*) |
| **Mcm2** | 99,2 | 7428555 | licensing factor | FGAQQDTIEVPEK (*2.2;2*); ESLVVNYEDLAAR (*3.5;2*); QLVAEQVTYR (*3.2;2*); VAVGELTDEDVK (*3.5;2*); DNNELLLFILK (*2.8;2*); VMLESFIDTQK (*2.7;2*); GLALALFGGEPK (*3.2;2*); DTVDPVQDEMLAR (*2.7;2*) |
| **Mcm7** | 95 | 33469968 | licensing factor | LFADAVQELLPQYK (*3.3;2*) |

**Table 1.** List of proteins identified in complex with e:H4. The columns correspond to their names, theoretical molecular weights, access numbers, known or annotated functions and the sequences of peptides identified by LC-MS/MS analysis. Given that many peptides have been identified several times, the highest confidence score (expressed as Xcorr values) for every particular peptide and the corresponding charge state are indicated in the brackets (score; charge).

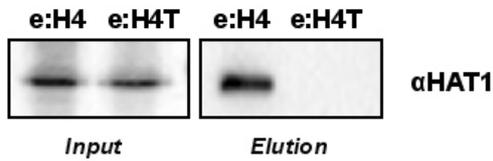

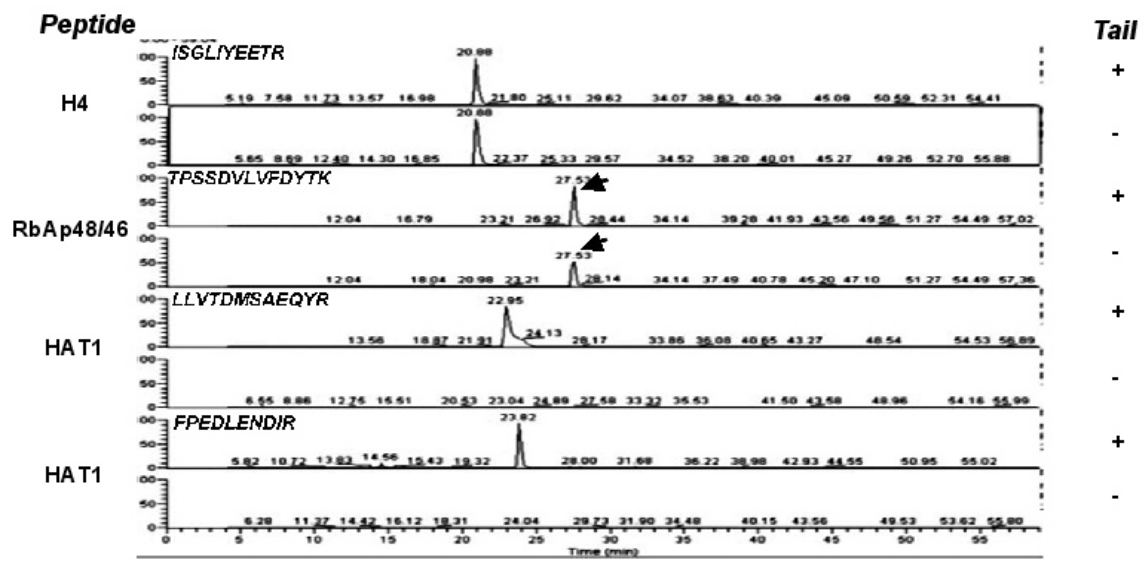

**Supplemental figure 1. HAT1 and RbAp48 are absent in H4T complex.** (A) Western blot analysis with anti-HAT1 antibody of the input (input) and the complex preparation sample (elution) from HeLa cells expressing e:H4 and tailless e:H4T. (B) MRM chromatogram showing presence of particular peptides in the previously described samples. The mass spectrometer was set to target specific peptides in order to confirm the presence of identified proteins by MS/MS analysis. '+' corresponds to the full length e:H4 and (-) corresponds to the tailless e:H4T complex. H4 peptide is detected in both e:H4 and e:H4T complexes. For the peptide common for RbAp48 and Rb46 proteins, the detected peak in decreased in the e:H4T complex (shown by arrows). HAT1 peptides (LLVTDMSAEQYR and FPEDLENDIR) are found in the e:H4 complex and absent in the e:H4T complex.